\documentclass[aps,prl,reprint,twocolumn,superscriptaddress,english,longbibliography]{revtex4-2}
\usepackage{amssymb}
\usepackage{graphicx}
\usepackage{amsmath}
\usepackage{epsfig}
\usepackage[latin9]{inputenc}
\usepackage{array}
\usepackage{multirow}
\usepackage{color}
\usepackage{esint}
\usepackage{bm}
\usepackage{bbm}
\usepackage{babel}
\usepackage{epstopdf}
\usepackage{mathtools}
\usepackage{soul}
\usepackage[dvipsnames]{xcolor}
\usepackage{tikz}
\usepackage[bookmarks=true,pdftex,colorlinks=true,urlcolor=blue,linkcolor=black,citecolor=blue,breaklinks=true,hypertexnames=false]{hyperref}
\usepackage{multibib} %right-align reference numbers
\newcommand*\chem[1]{\ensuremath{\mathrm{#1}}} % For Chemistry notation
\usepackage{siunitx} % SI Unit

\usepackage{float}

\usepackage{caption}

\DeclareCaptionLabelSeparator{bar}{\space\textbar\space}
\usepackage[labelfont=bf]{caption}
\addto\captionsenglish{\renewcommand{\figurename}{Fig.}}
\begin{document}

	\setcounter{equation}{0} \setcounter{figure}{0}
	\setcounter{table}{0} \setcounter{page}{1} \makeatletter
	\title{Observing the two-dimensional Bose glass in an optical quasicrystal}
	
	\author{Jr-Chiun Yu}
	\affiliation{Cavendish Laboratory, University of Cambridge, J.J. Thomson Avenue, Cambridge CB3 0HE, United Kingdom\looseness=-1}
 \author{Shaurya Bhave}
	\affiliation{Cavendish Laboratory, University of Cambridge, J.J. Thomson Avenue, Cambridge CB3 0HE, United Kingdom\looseness=-1}
 \author{Lee Reeve}
	\affiliation{Cavendish Laboratory, University of Cambridge, J.J. Thomson Avenue, Cambridge CB3 0HE, United Kingdom\looseness=-1}
  \author{Bo Song}
	\affiliation{Cavendish Laboratory, University of Cambridge, J.J. Thomson Avenue, Cambridge CB3 0HE, United Kingdom\looseness=-1}
  \author{Ulrich Schneider}
	\affiliation{Cavendish Laboratory, University of Cambridge, J.J. Thomson Avenue, Cambridge CB3 0HE, United Kingdom\looseness=-1}

    \begin{abstract}
        The combined effect of disorder and interactions is central to the richness of condensed matter physics and can lead to novel quantum states such as the Bose glass phase in disordered bosonic systems. Here, we report on the experimental realisation of the two-dimensional Bose glass using ultra-cold atoms in an eight-fold symmetric quasicrystalline optical lattice. By probing the  coherence properties of the system, we observe a Bose glass to superfluid transition and map out the phase diagram in the weakly interacting regime. Moreover, we reveal the non-ergodic nature of the Bose glass by probing the capability to restore coherence. Our observations are in good agreement with recent quantum Monte Carlo predictions and pave the way for experimentally testing the connection between the Bose glass, many-body localisation, and glassy dynamics more generally.
    \end{abstract}
    
    \maketitle
    \date{}

%=======================================================================================
\section{Introduction}

The interplay between disorder and interaction is central to the richness of condensed matter physics since any real-life material will inevitably contain a certain degree of impurities and defects, and inter-particle interactions are almost always present. While disorder tends to localise non-interacting particles, leading to Anderson localisation~\cite{anderson1958absence}, interactions can counteract this, resulting in conducting ergodic states. More generally, the combination of disorder and interactions gives rise to rich physics governed by reduced or absent relaxation and transport, such as glassy dynamics or non-ergodic many-body localised systems, and forms one of the central topics in quantum statistical physics during the last decade~\cite{abanin2019colloquium}.

\begin{figure}
\centering
\includegraphics[width=85mm]{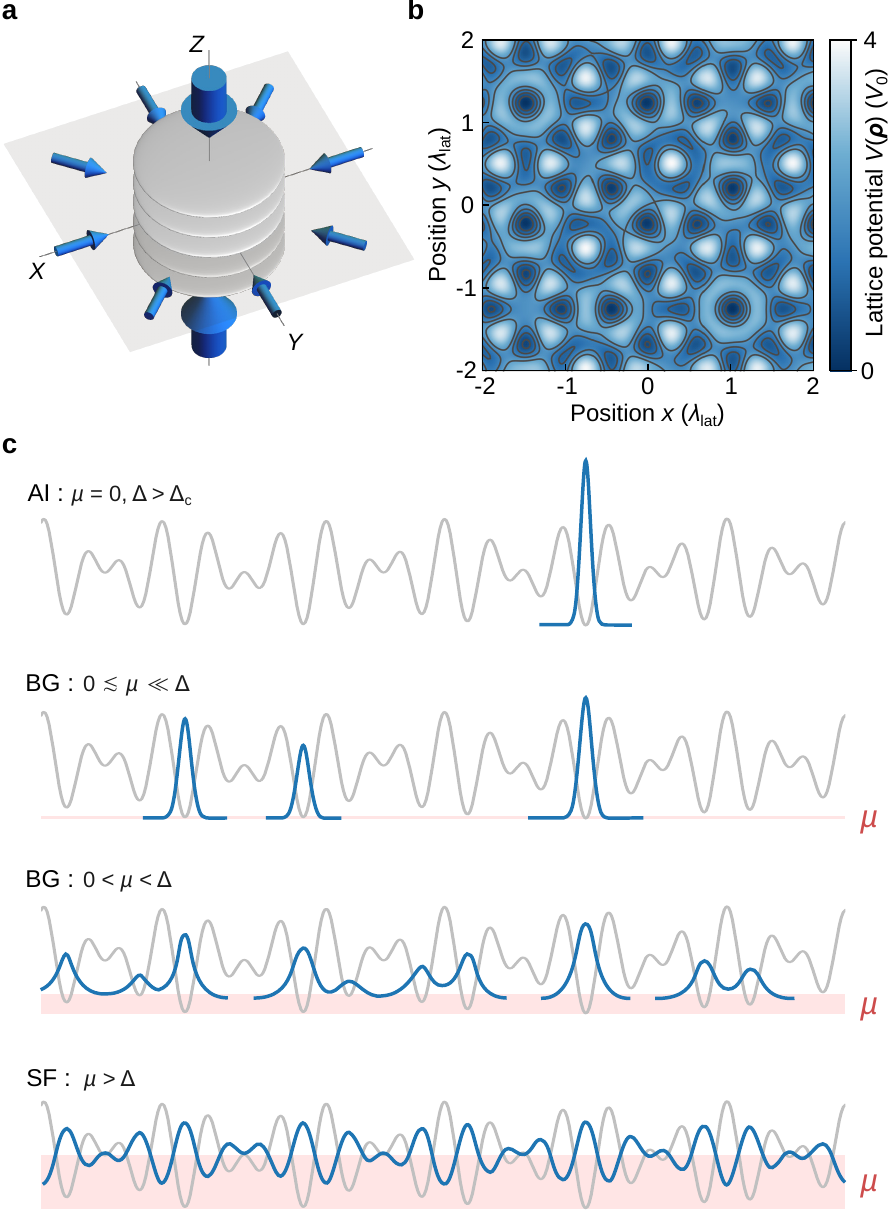}
\caption{\textbf{Lattice potential and sketch of possible phases.}
\textbf{a}, The 2D quasicrystalline optical lattice is generated by superimposing four independent 1D lattices in the $xy$-plane, marked by small arrows. A deep $z$ lattice (large arrows) divides the system into a series of independent quasi-2D layers. \textbf{b}, Examplary potential in a single layer. \textbf{c}, Repulsive interactions can delocalise an originally localised disordered system. Figures from top to bottom sketch the transition of the system's ground state with increasing chemical potential $\mu$, starting from the Anderson insulator (AI) in the non-interacting limit ($\mu=\epsilon_{0}=0$), the localised but compressible Bose glass (BG) for weak repulsive interactions where bosons spill over into other low-lying minima and form local superfluid puddles, and finally the superfluid (SF) when the chemical potential is comparable to the disorder strength $\Delta$.
}
\label{fig:QC_potential_test_figure}
\end{figure}

In bosonic systems, a hallmark of this interplay is the emergence of a novel ground-state phase, called Bose glass. 
The Bose glass is an insulating but compressible phase without long-range phase coherence~\cite{giamarchi1988anderson, fisher1989boson}. It was originally discussed purely as a ground state at zero temperature, but has been shown to extend to finite energy~\cite{michal2016finite, bertoli2018finite, zhu2022thermodynamic, ciardi2022finite}. In the weakly interacting regime, the Bose glass can be understood starting from a non-interacting Anderson insulator, where in the ground state all bosons localise at the lowest potential minimum; see Fig.\ref{fig:QC_potential_test_figure}c. 
Adding small repulsive interactions to such systems will lead to bosons spilling over into other low-lying orbitals in order to minimise the interaction energy. This regime has also been referred to as an Anderson glass or Lifshitz glass\,\cite{scalettar1991localization,lugan2007ultracold}. With increasing interactions, and thereby increasing chemical potential, these originally isolated orbitals will form local superfluid puddles that  will eventually merge into a global superfluid phase.

Since the lowest-lying minima will typically be located arbitrarily far away from each other, any changes to the system that require redistribution of particles between these will thus require arbitrarily long times, leading to non-ergodic behaviour of the Bose glass. In the non-interacting Anderson limit, orbitals localised at different local minima can indeed possess arbitrary close energies while having only exponentially weak couplings~\cite{altshuler2010anderson}, thus resulting in many almost degenerate levels. This absence of level repulsion is a hallmark of non-ergodic phases and has been shown numerically in a different context to extend to the many-body localised regime~\cite{oganesyan2007localization, pal2010many}. As a consequence of these exponentially small gaps, even slow parameter changes within the Bose glass will cause a significant number of excitations and take the system out of equilibrium. Therefore, the thermodynamic notion of quasi-static or adiabatic changes, where the system remains in thermal equilibrium at all times and the process is isentropic, does not apply. 
This unique feature indicates that the Bose glass is a localised, non-ergodic phase and opens the question to which degree it can be regarded as the low-energy limit of bosonic many-body localisation (MBL)~\cite{abanin2019colloquium}.

Disordered interacting bosons have been studied for instance using $^4$He in porous media~\cite{crowell1997}, Cooper pairs in superconducting films~\cite{sacepe2011localization}, and disordered quantum magnets~\cite{yu2012bose}. 
In the context of ultra-cold atoms, the Bose glass has been extensively studied using various numerical methods~\cite{rapsch1999density,  roux2008quasiperiodic,bissbort2009stochastic, soyler2011phase, niederle2015bosons, gerster2016superfluid, yao2020lieb,  zhang2015equilibrium, johnstone2021mean}. Initial experiments in one dimension demonstrated the loss of coherence but were strongly affected by finite-temperature effects~\cite{gadway2011glassy,fallani2007ultracold, deissler2010delocalization, d2014observation, gori2016finite} and experiments in three dimensions using speckle disorder studied momentum and quench responses~\cite{pasienski2010disordered, meldgin2016probing}.

In this work, we investigate the ground states of a weakly interacting Bose gas in a two-dimensional (2D) eight-fold rotationally symmetric quasicrystalline optical lattice~\cite{viebahn2019matter}. By analysing the momentum distribution of the system, we observe the Bose glass-to-superfluid phase transition, and map out the phase diagram in the weakly interacting regime. Furthermore, our work experimentally establishes the non-ergodic nature of the Bose glass, thereby highlighting its continuous connection to potential bosonic MBL phases at finite energy density~\cite{choi2016exploring,bordia2017probing}.
%
%---------------------------------------------
% Fig. 1: Lattice potential and sketch of possible phases
%---------------------------------------------
%------------------------------------------------
%

%===================================================================================================
\section{A 2D quasicrystalline optical lattice}

Quasicrystals are long-range ordered yet not periodic~\cite{shechtman1984metallic, steurer2018quasicrystals} and thereby represent a fascinating middle ground between order and disorder. In contrast to purely random potentials, where in one and two dimensions all single-particle eigenstates are localised for any non-vanishing disorder~\cite{abrahams1979scaling}, quasiperiodic potentials support a phase transition from extended to exponentially localised states at a finite potential depth~\cite{szabo2020mixed,roati2008anderson}, thus providing an ideal platform for studying disorder-induced phenomena.

In our experiment, we load a degenerate Bose gas of $\sim 1.2\times 10^{5}$ potassium $\chem{^{39}K}$ atoms into a 2D quasicrystalline optical lattice using a $\SI{45}{ms}$ long exponential ramp, see Methods. The optical quasicrystal is formed by superimposing four independent blue-detuned one-dimensional (1D) lattices in the $xy$-plane at $45^\circ$ angles, as depicted  schematically in Fig.\ref{fig:QC_potential_test_figure}a. Each of these lattices is a 1D standing wave created by a retro-reflected laser beam at wavelength $\lambda_{\text{lat}} = \SI{725.4}{nm}$. In addition, a deep lattice along the direction perpendicular to the plane ($z$-axis) effectively slices the system into an array of 2D layers (see the grey discs in Fig.\ref{fig:QC_potential_test_figure}a). The resulting potential is given by
%
%-----------------------------------------------
% Eq.1: Lattice potential 
%-----------------------------------------------
\begin{align}
    V(\bm{\rho}=\{x,&y\},z) = V_{0}\sum_{i=1}^{4}\sin^2(\mathbf{k_{i}} \cdot \bm{\rho} +\phi_{i})+V_{z}\sin^2(k_{z} z),\nonumber \\
    \mathbf{k_{i}}&\in \frac{2\pi}{\lambda_{\text{lat}}}
    \left\{
        \begin{pmatrix}
        1\\
        0\\
        \end{pmatrix},
        \frac{1}{\sqrt{2}}
        \begin{pmatrix}
        1\\
        1\\
        \end{pmatrix},
        \frac{1}{\sqrt{2}}
        \begin{pmatrix}
        -1\\
        1\\
        \end{pmatrix},
        \begin{pmatrix}
        0\\
        1\\
        \end{pmatrix}
        \right\}, 
\end{align}
%-----------------------------------------------
%
where $V_{0}$ and $V_{z}$ denote the lattice depths, and $\mathbf{k_{i}}$ and $k_{z}$ are the respective wave vectors ($\left|\mathbf{k_{i}}\right|=k_{z}=2\pi/\lambda_{\text{lat}}$) of the four 1D lattices in the $xy$-plane and the $z$ lattice. The phase offsets $\phi_{i}$ are central to describe phasonic degrees of freedom and topological pumping in these potentials, but play no significant role for localisation in large systems~\cite{gottlob2022hubbard}.

Throughout this work, the depths of the horizontal lattices are varied in the range of $V_{0}=1$--$4\,E_{\text{rec}}$ while the $z$ lattice is kept at $V_{z}=20$\,$E_{\text{rec}}$, where $E_{\text{rec}} = \hbar^{2} k_{\text{lat}}^{2}/(2m)$ is the recoil energy, $\hbar$ is the reduced Planck constant and $m$ is the atomic mass. The deep $z$ lattice provides a sufficiently strong vertical confinement so that inter-layer tunnelling is negligible. As a consequence, atoms loaded into the lattice will be tightly confined to individual quasi-2D systems that exhibit an eight-fold symmetric quasicrystalline structure, as depicted in  Fig.\ref{fig:QC_potential_test_figure}b. 

Even though the lattice depths used for the 2D quasicrystalline lattice are rather low, the physics of the system is nonetheless captured by a dedicated quasiperiodic Bose-Hubbard model~\cite{gottlob2022hubbard}, which in second quantisation reads
%
%-----------------------------------------------
% Eq2: Quasiperiodic Bose-Hubbard model
%-----------------------------------------------
\begin{align}
\hat{H}_{\text{QBH}}&=\sum_{i}\epsilon_{i}\hat{a}^{\dagger}_{i}\hat{a}_{i} - \sum_{i\neq j} J_{ij}\hat{a}^{\dagger}_{i}\hat{a}_{j}+\sum_{i}\frac{U_{i}}{2}\hat{n}_{i}(\hat{n}_{i}-1).
\label{Disordered_Bose_Hubbard_model}
\end{align} 
%------------------------------------------------
%
%
%------------------------------------------------
% Fig2: Bose glass to superfluid transition
%------------------------------------------------
\begin{figure*}
\centering
\includegraphics[width=180mm]{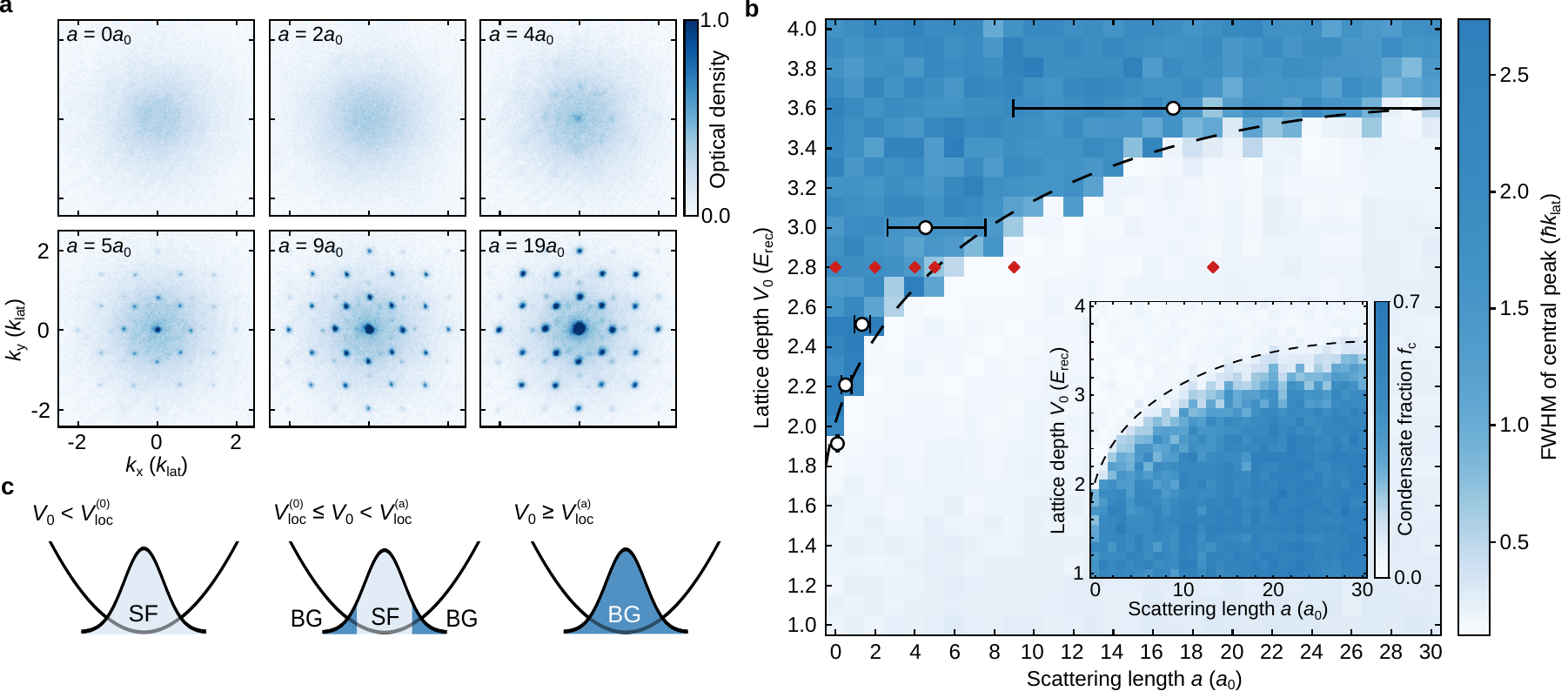} 
\caption{\textbf{Bose glass to superfluid transition.} \textbf{a}, Time-of-flight images ($\SI{9}{ms}$ TOF, $5$ shots averaged) for different scattering lengths $a$ at a fixed lattice depth of $V_0=2.8\,E_{\text{rec}}$. While the system is localised in the non-interacting and very weakly interacting cases, the appearance of sharp interference peaks for stronger interactions signals the emergence of long-range phase coherence, characteristic for the superfluid. \textbf{b}, Width of the central peak, distinguishing the coherent superfluid (light blue) from the incoherent Bose glass (dark blue). The dashed line is a guide to the eye indicating the detected phase boundary in the centre of the cloud $V_{\text{loc}}^{(a)}$. It is identical to the line shown in the inset and in Fig.\ref{fig:Restoring_coherence}d. White points and error bars denote the QMC prediction from Ref.~\cite{gautier2021strongly}, see Methods. Images in panel `a' correspond to parameters marked by red diamonds. The inset shows the condensate fraction $f_{c}$ extracted from the same set of images, highlighting the coexistence of the two phases. \textbf{c}, Phase transition in inhomogeneous system. The shaded Gaussian denotes the in-trap atomic density and the parabola represents the external trapping potential. For shallow lattices, the ground state is purely superfluid (left). At the non-interacting critical depth $V_{\text{loc}}^{(0)}$, the Bose glass starts to appear at the low-density edge of the cloud where interaction effects are small (middle). With increasing lattice depth, the phase boundary gradually moves inwards until the entire cloud enters the Bose glass phase at $V_{\text{loc}}^{(a)}$ (right).}
\label{fig:phase_diagram_test_figure}
\end{figure*}
%------------------------------------------------
%
Here $\hat{a}^{\dagger}_{i}$ ($\hat{a}_{i}$) is the bosonic creation (annihilation) operator  on the \textit{i}th lattice site, and $\hat{n}_{i}=\hat{a}^{\dagger}_{i}\hat{a}_{i}$ is the corresponding number operator. The Hamiltonian $\hat{H}_{\text{QBH}}$ is characterised by three site-dependent parameters, namely on-site energies $\epsilon_{i}$, tunnelling energies $J_{ij}$, and on-site interactions $U_{i}\propto a$, whose scale can be independently controlled by tuning the atomic s-wave scattering length $a$ by means of a Feshbach resonance, see Methods. 
 We set $\epsilon_0\vcentcolon=\min{\epsilon_i}=0$ and use $\Delta\vcentcolon=\max{\epsilon_i}$ as an intuitive measure of ``disorder strength", even though the modulation in $J_{ij}$ and $U_{i}$ also influences the physics. 

In the weakly interacting regime, systems described by the Hamiltonian $\hat{H}_{\text{QBH}}$ host a phase transition from Bose glass to superfluid, as illustrated in Fig.\ref{fig:QC_potential_test_figure}c. At strong interactions with $U\gg J$, they furthermore host commensurate Mott insulators~\cite{gottlob2022hubbard,gautier2021strongly}, however, this regime is not probed in the current paper, see Methods.
In this strongly interacting regime, the term Bose glass was introduced to describe the phase emerging when the charge order of the Mott insulator vanishes for strong enough disorder $\Delta \approx U$~\cite{soyler2011phase, zhang2015equilibrium}. This regime exhibits the same phenomenology as the weakly interacting Bose glass, namely being a compressible, gapless, insulating phase without long-range coherence, and hence they both belong to the same Bose glass phase.

%===================================================================================================
\section{Phase diagram}

Our main observable to distinguish superfluid and localised states is the momentum distribution detected using time-of-flight (TOF) imaging, i.e., by releasing the atomic cloud from all trapping potentials and imaging its density distribution after $9\,$ms of free expansion. This can be understood as a matter-wave diffraction experiment where waves originating on different lattice sites expand, overlap, and then interfere. Analogous to diffraction experiments in optics and in periodic lattices~\cite{pedri2001expansion, greiner2002quantum}, the coherence length, i.e., the range of spatial coherence between lattice sites, determines the width of the matter-wave interference peaks. A high-contrast interference pattern composed of sharp peaks indicates the presence of long-range phase coherence, the signature of the superfluid phase. 
Localised states with only short-range coherence, on the other hand, result in an incoherent broad momentum distribution.

Fig.\ref{fig:phase_diagram_test_figure}a presents a series of TOF images recorded for different scattering lengths at a fixed lattice depth of $V_{0}=2.8$\,$E_{\text{rec}}$. At this lattice depth, the single-particle ground state is strongly localised~\cite{sbroscia2020observing}, and the measured momentum distribution at  vanishing scattering length (top-left panel) correspondingly exhibits the broad momentum profile of a localised Anderson insulator. With increasing interactions, however, we observe a clear phase transition from the incoherent Bose glass  to a superfluid with sharp, high-contrast interference peaks.

In order to quantitatively study this transition, we fit the central peak of each individual TOF image using 2D Gaussians and extract its full width at half maximum (FWHM) as a measure of the coherence length. The resulting phase diagram is shown in Fig.\ref{fig:phase_diagram_test_figure}b and clearly reveals two distinct phases: the coherent superfluid at shallow lattices (light blue) turns abruptly into the incoherent Bose glass (dark blue) at an interaction-dependent critical lattice depth $V_{\text{loc}}^{(a)}$. 
At vanishing scattering length, the observed $V_{\text{loc}}^{(0)}$ coincides with the known single-particle localisation point at around $V_{\text{loc}}^{(0)}=1.78(2)$\,$E_{\text{rec}}$~\cite{szabo2020mixed,sbroscia2020observing} up to minor corrections ($\lesssim 1\, a_0$) stemming from the presence of weak residual interactions due to the small dipole-dipole interactions~\cite{wall2013dipole} and calibration uncertainties, see Methods. With increasing scattering lengths, the critical lattice depth $V_{\text{loc}}^{(a)}$ indicated by the dashed line shifts considerably towards deeper lattices, directly demonstrating that even weak repulsive interactions can significantly counteract localisation. The observed transition agrees well with the recent quantum Monte Carlo (QMC) simulations reported in Ref.~\cite{gautier2021strongly}.

As a complementary observable, the inset of Fig.\ref{fig:phase_diagram_test_figure}b shows the same phase diagram analysed in terms of the condensate fraction $f_{c}\vcentcolon=\mathcal{N}_{\text{coh}}/\mathcal{N}$, i.e., the number of atoms in the sharp interference peaks  $\mathcal{N}_{\text{coh}}$ divided by the total atom number $\mathcal{N} = \mathcal{N}_{\text{coh}}+\mathcal{N}_{\text{incoh}}$, where $\mathcal{N}_{\text{incoh}}$ represents the population of the incoherent background, see Methods for details. The condensate fraction is high for shallow lattices and begins to decrease slowly after the lattice depth exceeds the non-interacting critical depth $V_{\text{loc}}^{(0)}$; see also Fig.\ref{fig:Restoring_coherence}e. This initially small downward trend gradually becomes stronger, and the condensate fraction eventually reaches zero at the same critical depth $V_{\text{loc}}^{(a)}$ extracted from the FWHM measurement (dashed line).

The gradual decrease in the condensate fractions implies the coexistence of superfluid and Bose glass in the system. This is the result of the inhomogeneous atomic density caused by the background harmonic dipole trap, as illustrated in Fig.\ref{fig:phase_diagram_test_figure}c: when atoms are loaded into the lattice, the low-density edge of the cloud, where interaction effects vanish, will start to localise at the critical depth for non-interacting atoms $V_{\text{loc}}^{(0)}$~\cite{meldgin2016probing}. As we further increase the lattice depth, the phase boundary that separates the Bose glass from the superfluid core will slowly move towards higher densities until all atoms are ultimately in the Bose glass phase.

While the condensate fraction provides trap-averaged information, the almost binary signature provided by the FWHM shown in the main diagram of  Fig.\ref{fig:phase_diagram_test_figure}b is sensitive to the phase transition in the centre of the trap, i.e., it describes the point when atoms in the centre of the trap localise. This is because the observed widths of all superfluid peaks are dominated by the finite initial cloud size such that no obvious broadening can be detected until all peaks have completely merged into the incoherent background.

%%==================================================================================================
\section{Non-ergodic nature of the Bose glass}

In typical quantum phase transitions between ergodic phases, for example from  superfluid to Mott insulator~\cite{greiner2002quantum}, an important experimental check is whether the phase transition was crossed adiabatically, and thereby reversibly, or whether the observed loss of coherence results from irreversible heating, for instance due to rapid non-adiabatic changes that generate entropy. In the present case, however, the situation is expected to be rather different, as the Bose glass is non-ergodic and the thermodynamic notion of adiabatic changes does not apply. 

\begin{figure*}
\centering
\includegraphics[width=180mm]{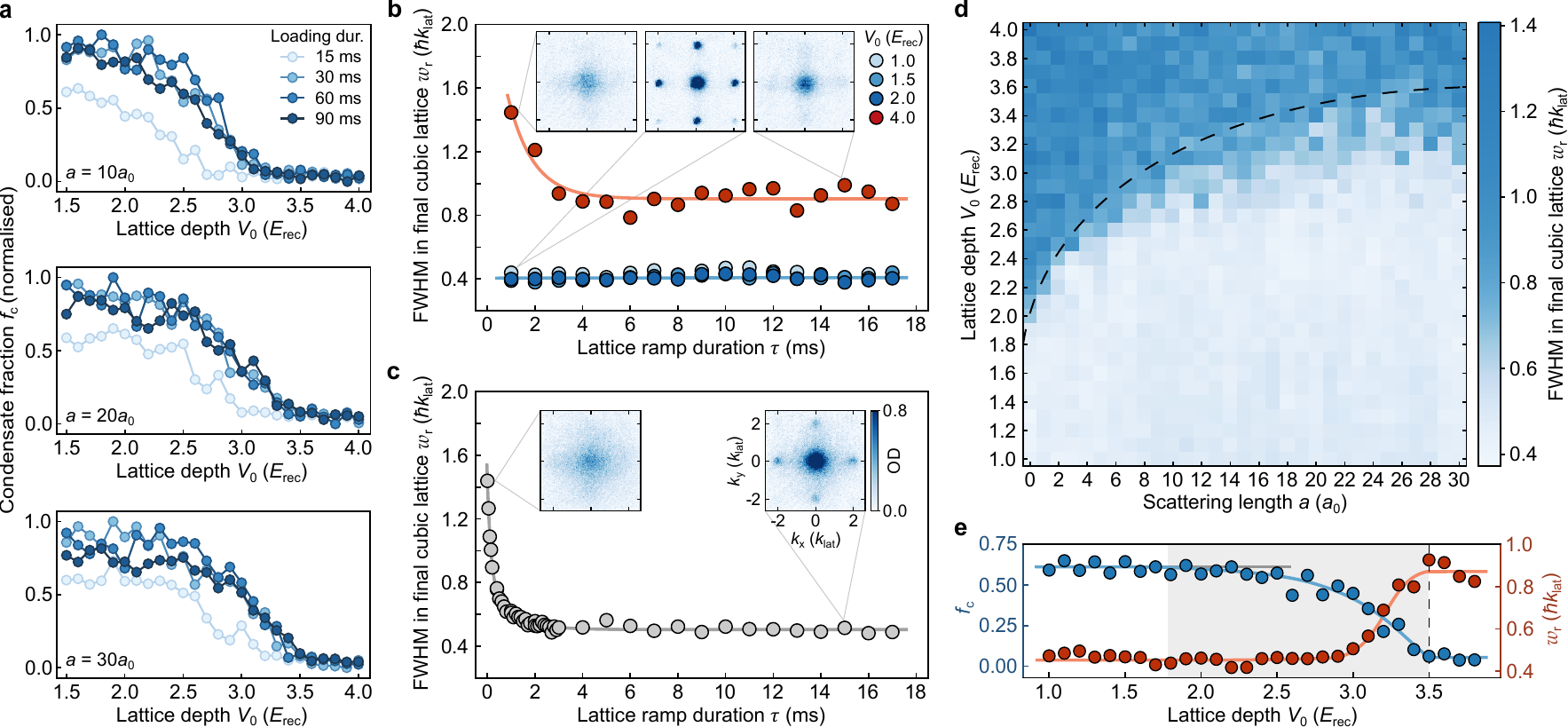}
\caption{\textbf{Non-ergodicity of the Bose glass.} \textbf{a}, Condensate fraction in 2D quasicrystal (normalised within each plot) as a function of lattice depth for different loading durations and scattering lengths. While $\SI{15}{ms}$ ramps result in consistently lower condensate fractions, there is no consistent difference between $\SI{30}{ms}$ and longer ramps. \textbf{b}, FWHM of the central peak ($w_{r}$) after a linear ramp of duration $\tau$ from the 2D quasicrystal into a regular 3D cubic lattice, where the ground state is a superfluid. Coloured circles correspond to different depths of the quasicrystalline potential $V_0$ for a fixed scattering length of $a=10\,a_{0}$. For $V_{0}<V_{\text{loc}}^{(10)}$ (blue circles),  the initial state in the quasicrystal is superfluid and the final states show strong superfluid order for all explored ramp times. For a deep Bose glass at $V_{0}>V_{\text{loc}}^{(10)}$ (red circles), in contrast, there is no initial coherence and only a very limited degree of phase coherence can be restored. \textbf{c}, An equivalent measurement for a Mott insulator in a regular 3D cubic lattice ($V_{x,y,z}=16\,E_{\text{rec}}$, $a=150\,a_{0}$). While the initial state also lacks coherence, it can be rapidly restored by ramping down the lattice depth in $\tau\gtrsim \SI{2}{ms}$. \textbf{d}, Phase diagram showing $w_{r}$ for a slow ramp with $\tau=\SI{15}{ms}$ highlighting three different regimes: a pure superfluid (SF) (light blue), an intermediate regime where superfluid and Bose glass coexist in the trap, and finally the pure Bose glass (darker blues). The transition into the pure Bose glass is consistent with the phase boundary extracted in Fig.\ref{fig:phase_diagram_test_figure}b (dashed line). \textbf{e}, Comparing condensate fraction $f_c$ in the quasicrystal with  $w_r$ for $a=23\,a_0$, demonstrating the consistency of all observations. Dashed line denotes the critical lattice depth $V_{\text{loc}}^{(23)}$ extracted from the main diagram of  Fig.\ref{fig:phase_diagram_test_figure}b and the grey area indicates the intermediate regime where superfluid and Bose glass coexist. Solid lines in b,c,e are guides to the eye.}
\label{fig:Restoring_coherence}
\end{figure*}

To demonstrate this, we first study in Fig.\ref{fig:Restoring_coherence}a the effect of different lattice loading durations on the resulting condensate fraction. A too rapid lattice ramp ($\SI{15}{ms}$) gives rise to considerable heating already in the superfluid regime, leading to lower condensate fractions compared to slower ramps. Once the loading duration exceeds $\SI{30}{ms}$, in contrast, the condensate fraction becomes independent of the loading rate, and all measurements reveal a consistent critical lattice depth $V_{\text{loc}}^{(a)}$ that for a given density depends solely on the interaction strength. 

Despite the loading duration clearly becoming irrelevant for sufficiently slow lattice ramps, the non-ergodic nature of the Bose glass crucially implies that once the system has entered the Bose glass, it cannot be transformed back into a superfluid, as the excitations generated inside the non-ergodic regime result in a significant entropy increase. We demonstrate this defining feature by first loading the atoms into the 2D quasicrystalline lattice in $\SI{45}{ms}$ before continuously transforming the non-periodic lattice into a periodic simple-cubic lattice, where the ground state is a superfluid for all studied parameters. This transformation is carried out by linearly ramping the depth of the $x$, $y$, and $z$ lattices to $8\,E_{\text{rec}}$ over various durations $\tau$ while simultaneously reducing the depth of the remaining two diagonal lattices (see Fig.\ref{fig:QC_potential_test_figure}a) to zero.

%
%------------------------------------------------
% Fig.3: Non-ergodicity of the Bose glass
%------------------------------------------------
%
%------------------------------------------------

Fig.\ref{fig:Restoring_coherence}b presents the FWHM of the central  peak, $w_{r}$, in the final periodic lattice for different ramp times $\tau$ at a fixed scattering length ($a=10\,a_{0}$), and the outcome highlights the fundamentally distinct behaviours of the superfluid and Bose glass phases. For $V_{0}<V_{\text{loc}}^{(10)}$ (blue circles), the system remained superfluid during the entire sequence, and the ground state can adapt rapidly from a quasiperiodic extended wave to a periodic Bloch wave, as indicated by the sharp and narrow diffraction peaks for all ramp durations. For  $V_{0}>V_{\text{loc}}^{(10)}$ (red circles), however, where the system has entered the Bose glass regime, the initial state only contains very short range coherence and hence results in a high $w_{r}$. Furthermore, the measured $w_{r}$ remains significantly above that of the superfluid even for the slowest ramps explored in this measurement. This demonstrates that systems in this regime can only restore a very limited degree of phase coherence and thereby directly highlights the significant entropy production arising from the non-adiabatic evolution within the Bose glass.

In order to demonstrate that the reduced coherence is not caused by dynamical effects such as Kibble-Zurek-type dynamics~\cite{braun2015emergence} during too-fast final ramps, Fig.\ref{fig:Restoring_coherence}c shows an equivalent measurement starting from a Mott insulator in a deep three-dimensional (3D) simple-cubic lattice, where phase coherence is recovered by reducing the lattice potential to the same final depth as in the previous case. In this case, sharp interference patterns can be recovered already within $\SI{2}{ms}$ of ramp-down time, consistent with previous observations~\cite{greiner2002quantum, braun2015emergence}. 
This contrast not only experimentally confirms that the incoherent localised phase we observe in the optical quasicrystal is distinct from a Mott insulator, but also directly establishes the non-ergodic nature of the Bose glass.

Making use of this distinctive feature of the Bose glass, we mapped out another phase diagram. Fig.\ref{fig:Restoring_coherence}d shows the FWHM of the central peak ($w_{r}$) after a slow final ramp of $\tau=\SI{15}{ms}$ as a function of scattering length and depth of the intermediate quasicrystalline lattice. The dashed line is identical to the one in Fig.\ref{fig:phase_diagram_test_figure}b, indicating the phase transition in the cloud centre. This demonstrates the consistency of the different measurements: as the atoms localise and enter the Bose glass, not only does the condensate fraction decrease but also the coherence cannot be restored; see also Fig.\ref{fig:Restoring_coherence}e.

%===================================================================================================
\section{Conclusion}

In this work, we experimentally study the 2D Bose glass in an  optical quasicrystal with eight-fold rotational symmetry by probing the coherence properties of the system. We directly observe the phase transition between the Bose glass and the superfluid, in good agreement with quantum Monte Carlo simulations~\cite{gautier2021strongly}. In addition, we experimentally establish the non-ergodic character of the Bose glass by probing the capability to restore coherence. This paves the way for experimentally testing the connection between the Bose glass, many-body localisation (MBL), and glassy dynamics more generally. 
Quasicrystalline and quasiperiodic lattices offer a unique route to study MBL, as their long-range ordered nature can exclude conventional ergodic rare regions~\cite{szabo2020mixed, vstrkalj2022coexistence} that are expected to destabilise MBL by seeding thermalisation avalanches in real random systems~\cite{de2017stability,leonard2023probing}. 

%===================================================================================================
\bibliography{BoseGlass.bib}

\newpage

%===================================================================================================
\section{Methods}

%===================================================================================================
\subsection{Experimental sequence}

The experimental sequence begins with loading a Bose-Einstein condensate (BEC) of $\sim 1.2 \times 10^{5}$ \chem{^{39}K} atoms from a red-detuned crossed optical dipole trap ($\lambda_{\text{dip}}= \SI{1064}{nm}$, $(w_x,\,w_y,\,w_z)=2\pi\cdot(55,\,43,\,330)$\,Hz) into a blue-detuned 2D quasiperiodic optical lattice ($\lambda_{\text{lat}}= \SI{725.4}{nm}$). During the loading, the individual lattice depths are increased in $\SI{45}{ms}$ from zero to their target values using exponential ramps with a time constant of $\SI{10}{ms}$. The used target depths for the four horizontal lattices range within $V_{0}=1$\,--\,$4$\,$E_{\text{rec}}$ while a fixed depth of $V_{z}=20$\,$E_{\text{rec}}$ for the vertical $z$ lattice ensures the formation of well-defined quasi-2D systems. After this ramp, the atoms are held in the quasicrystal for $\SI{10}{ms}$. For imaging, we apply a short ``\textit{booster stage}''~\cite{stoferle2004transition} before we switch off all trapping potentials and record the matter-wave interference pattern by taking an absorption image after $\SI{9}{ms}$ time-of-flight (TOF).

The booster stage consists of linearly increasing the potential depth of the horizontal lattices in $40\,\mu$s to a final depth of $V_{\text{final}}=6\,E_{\text{rec}}$. 
This stage is sufficiently short to not change the coherence properties of the system while providing a tighter on-site confinement and thereby not only enhancing the brightness of high-order diffraction peaks but also significantly reducing the heavy saturation on the central momentum peak; see Extended Data Fig.\ref{fig:Booster_stage}a,\,b. 

The interaction strengths $U_i\propto a$ are independently controlled by tuning the atomic s-wave scattering length ($a$) using the Feshbach resonance close to $\SI{403}{G}$ of the $\left|F=1,m_F=1\right>$ state in \chem{^{39}K}~\cite{d2007feshbach, fletcher2017two}.  To ensure broadly comparable density distributions, the scattering length is initially prepared at a common finite value of $a=12\,a_{0}$ before the lattice loading starts and is then linearly ramped to the desired value within $a=0$\,--\,$30$\,$a_{0}$  during the last $\SI{20}{ms}$ of the lattice ramp. Subsequently, the scattering length remains constant until being suddenly switched to $a=0$\,$a_{0}$ at the beginning of the time-of-flight.

The initial Mott insulating state in Fig.\ref{fig:Restoring_coherence}c is prepared in a regular 3D lattice of depth $V_{0}=16\,E_{\text{rec}}$ at scattering length  $a=150\,a_{0}$. The restoration of phase coherence is then carried out by employing a $16$--$8\,E_{\text{rec}}$ linear ramp on all the three lattice axes simultaneously.

%===================================================================================================
\subsection{Extraction of condensate fraction}

The condensate fraction  $f_{c}$ is evaluated for every shot according to $f_{c} = \mathcal{N}_{\text{coh}}/\mathcal{N},$
where $\mathcal{N}_{\text{coh}}$ is the population in the sharp interference peaks, and $\mathcal{N}=\mathcal{N}_{\text{coh}}+\mathcal{N}_{\text{incoh}}$ is the total atom number with $\mathcal{N}_{\text{incoh}}$ 
being the number of atoms in the incoherent background.

To extract $\mathcal{N}_{\text{coh}}=\sum_{k}n_{k}$  from the time-of-flight (TOF) images, we first identify the most pronounced $81$ diffraction peaks within the first six diffraction orders~\cite{viebahn2019matter}  and then extract their populations $n_{k}$  by fitting independent 2D Gaussian profiles to each peak. To prevent counting spurious populations from weakly populated peaks, we exclude fitted populations $n_{k}$ below $0.12\%$ of the total atom number. 
Extended Data Fig.\ref{fig:Booster_stage}c illustrates the extracted populations.  
We note that the width of the diffraction peaks in the TOF images is dominated by the finite initial cloud size~\cite{braun2015emergence}. Therefore, no obvious broadening can be detected when the inhomogeneous system enters the Bose glass until all the interference peaks have completely merged into the incoherent background. 

The atom number in the incoherent background, $\mathcal{N}_{\text{incoh}}$, is acquired by performing an additional 2D Gaussian fit to whole cloud (region of interest $3.3\times3.3\,(\hbar k_{\text{lat}})^{2}$), where all detected diffraction peaks were masked during the fitting.

%===================================================================================================
\subsection{Parameter Calibration}

The two main experimental parameters are the lattice depth and the scattering length between atoms. The lattice depth is calibrated to within $0.1\,E_{\text{rec}}$ by analysing the dynamics of Kapitza Dirac diffraction for each 1D lattice individually; see the supplemental material of  Ref.\,\cite{viebahn2019matter} for details. 

The scattering length is calibrated by observing the prominent atom-loss features corresponding to the zero-crossing of the scattering length, where the in-situ density is highest, and the Feshbach resonance, where the loss coefficient is maximal.  We then interpolate the scattering length  between them using the common functional form~\cite{fletcher2017two, eigen2016observation}. As an independent crosscheck, the magnetic field is calibrated using radio frequency spectroscopy of the $\left|F=1,m_{F}=-1\right>$ to $\left|F=1,m_{F}=0\right>$ transition in rubidium and converted to a scattering length using literature values for the parameters of the Feshbach resonance~\cite{fletcher2017two, eigen2016observation}. The two approaches agree to  $\lesssim 1\, a_0$. 

%===================================================================================================
\subsection{Comparing to quantum Monte Carlo simulations}

The Quantum Monte Carlo  calculations reported in Ref.~\cite{gautier2021strongly} were performed as a function of the density $n$ of a homogeneous system at fixed interaction strength $g$. Since the main panel of Fig.\ref{fig:phase_diagram_test_figure}b focuses on the phase transition in the centre of the trap, we extract the experimental central density $n_0$ from in-situ absorption images using the known aspect ratio of the trap. In order to minimise statistical noise, we measure $n_0$ at different scattering lengths ($a=0$\,--\,$30\,a_{0}$) and constant lattice depth ($V_{0}=1 \,E_{\text{rec}}$) and find a mild interaction dependence $n_0(30a_0)\approx1/2\, n_0(0a_0)$ for the used lattice ramp. In addition, we relate the 2D interaction coupling constant $g$ used in \cite{gautier2021strongly} back to the 3D scattering length $a$ via
%
%------------------------------------------------
% Eq: Conversion formulas for scattering length
%------------------------------------------------
\begin{align}
g=\frac{\hbar^{2}}{m}\tilde{g}&\hspace{1mm}, \hspace{2mm}\tilde{g}\approx \tilde{g}_{0}=\frac{2\pi}{\ln\left(a_{\text{lat}}/a_{\text{2D}}\right)}\hspace{1mm}, \nonumber \\
a_{\text{2D}}=&\,2.092\,l_{\perp} \exp\left(-\sqrt{\frac{\pi}{2}}\frac{l_{\perp}}{a}\right). \nonumber
\end{align} 
%------------------------------------------------
%
 Here,  $a_{\text{lat}}=\lambda_{\text{lat}}/2$ and  $l_{\perp}= \sqrt{\hbar/m\omega_{\perp}}$ is the characteristic confining length given by the strong $z$ lattice with a trapping frequency of  $\omega_{\perp}=2\pi \cdot 87$\,kHz.

%===================================================================================================
\subsection{Excluding Mott insulators}

In order to investigate the possibility of Mott insulators in our experiment, we numerically compute the Bose-Hubbard parameters of the quasiperiodic potential using the results from Ref.~\cite{gottlob2022hubbard}. We calculate the site-dependent ratio between on-site interactions and tunneling energies $U_{i}/\sum_{j} \left|J_{ij}\right|$, where the sum runs over all significant tunnelling elements linking site $i$ to other adjacent sites. Within the explored parameter regime, this ratio reaches a maximum of $\max\left(U_{i}/\sum_{j} \left|J_{ij}\right|\right) \approx 1.4$ for the case of $a=30\,a_{0}$ and $V_{0}=4.0\,E_{\text{rec}}$.  This is significantly below the critical interaction strength for forming a Mott insulator in a 2D square lattice $\left(U/zJ\right)_{c} \approx 4.385$~\cite{wessel2004quantum}, where $z=4$ represent the number of the nearest neighbours. Furthermore, the studied parameter range lies within the weakly interacting regime of Ref.~\cite{zhu2022thermodynamic}, and Mott insulators can hence be excluded in this experiment.

\vspace{5mm}

%===================================================================================================
\textbf{Acknowledgements}
We would like to thank Konrad Viebahn, Matteo Sbroscia and Edward Carter for their contributions to building the experimental setup and are grateful to Emmanuel Gottlob, Joseph Thywissen, and Laurent Sanchez-Palencia and his team for discussions. This work was supported by the European Commission ERC Starting Grant QUASICRYSTAL, the EPSRC Grant (No.\,EP/R044627/1), and EPSRC Programme Grant DesOEQ (No.\,EP/P009565/1).

\vspace*{5mm}
\newpage

%===================================================================================================
\subsection{Extended Data}
\setcounter{figure}{0}
\renewcommand{\figurename}{Extended Data Fig.}
\renewcommand{\thefigure}{\arabic{figure}}
%
%------------------------------------------------
% Extended Data Fig.1 : Effect of  booster stage and an example of the population extraction.
%------------------------------------------------
%
\begin{figure}[h]
\centering
\includegraphics[width=85mm]{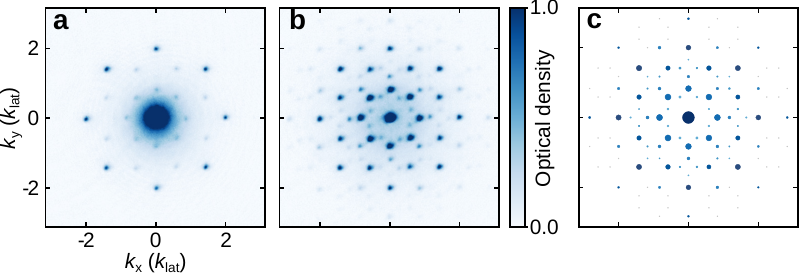}
\caption{\textbf{Effect of  booster stage ($V_{0}=2\,E_{\text{rec}}$, $a=10\,a_{0}$) and an example of the population extraction.} \textbf{a}, In the absence of the booster, the majority of condensed atoms remain in the central diffraction peak, with only a small fraction occupying the satellite peaks. The high atomic density of the central peak causes almost all the imaging light around this central area to be fully absorbed, leading to significant imaging saturation at $\mathbf{k}=0$. \textbf{b}, the booster stage promotes condensed atoms to higher diffraction orders, thus facilitating the fitting. \textbf{c},  Simulated diffraction pattern for the first $6$ diffraction orders. The $81$ peaks considered in the population count are coloured in blue, with their radius indicating the extracted population $n_{k}$. Gray dots represent the peaks that can also be observed but are not included in the population count due to their low populations. Images  in panels a \& b are averaged over $30$ experimental shots in order to visually emphasise the signal from very weakly populated high-order diffraction peaks.}
\label{fig:Booster_stage}
\end{figure}
%------------------------------------------------
%

\clearpage

\end{document}